\documentclass[twocolumn,showpacs,showkeys]{revtex4}
\usepackage{amssymb}
\usepackage{amsmath}
\usepackage{graphicx}

\begin{document}

\title{Dynamics of quasi-stationary systems: Finance as an example}

\author{Philip Rinn$^1$}
\email{philip.rinn@uni-oldenburg.de}

\author{Yuriy Stepanov$^2$}

\author{Joachim Peinke$^1$}

\author{Thomas Guhr$^2$}

\author{Rudi Sch{\"a}fer$^2$}
\affiliation{
$^1$ Institute of Physics and ForWind, Carl-von-Ossietzky University Oldenburg, Germany\\
$^2$ Faculty of Physics, University of Duisburg-Essen,  Duisburg, Germany}



\begin{abstract}
We propose a combination of cluster analysis and stochastic process analysis to characterize high-dimensional complex dynamical systems by few dominating variables. As an example, stock market data are analyzed for which the dynamical stability as well as transitions between different stable states are found. This combined method also allows to set up new criteria for merging clusters to simplify the complexity of the system. The low-dimensional approach allows to recover the high-dimensional fixed points of the system by means of an optimization procedure.
\end{abstract}

\maketitle

\section{Introduction}
For complex dynamical systems consisting of many interacting subsystems it is a general challenge to reduce the high dimensionality to a few dominating variables that characterize the system. Cluster analysis is a method to group elements according to their similarity. However, there is an ambiguity as different algorithms may lead to different clusters \cite{Estivill-Castro2002, Jain2010}.
The dynamical behavior of complex systems may reduce their complexity by self-organi\-zation. Here the high-dimensional dynamics generates a few order parameters evolving slowly on a strongly fluctuating background \cite{Haken2004}. With the help of stochastic methods it is possible to show such simplified dynamics and to estimate a Langevin equation directly from the data, \textit{i.e.} the data are analyzed as a stochastic process with drift and diffusion term \cite{Friedrich2011}.

We show that cluster analysis and stochastic methods can be combined in a fruitful way. Cluster analysis does not aim at grasping dynamical effects. By combining it with stochastic methods we show that an improved dynamical cluster classification can be obtained. Furthermore we extract dynamical cluster features of a complex system as emerging and disappearing clusters. Related ideas were put forward by Hutt \textit{et al.} \cite{Hutt2000} for detecting fixed points in spatiotemporal signals. Here, we focus on financial data, for which quasi-stationary market states were identified in the correlation structure using cluster analysis \cite{Muennix2012}. We briefly sketch these findings before developing our combined analysis.

\section{Market States: Clustering}
We study the daily closing prices $S_k(t)$ of the S\&P 500 stocks for the period 1992 -- 2012 \cite{yahoo2013}, which aggregate to 5189 trading days. Only the 306 stocks that were continuously traded during the whole period are considered. We calculate the relative price changes, \textit{i.e.} the returns

  \begin{equation}
    r_k(t) = \frac{S_k(t + \Delta t) - S_k(t)}{S_k(t)} ,
    \label{eq.ret}
  \end{equation}

\noindent for each stock $k$ for a time horizon $\Delta t$ of one trading day (td). To avoid an estimation bias due to time-varying trends and fluctuations, we perform a local normalization \cite{Schaefer2010} of the returns before estimating their correlations. We denote the locally normalized returns by $\tilde{r}_k(t)$. We then calculate the Pearson correlation coefficients

  \begin{equation}
    C_{ij} = \frac{\langle \tilde{r}_{i}\tilde{r}_{j}\rangle -\langle \tilde{r}_{i}\rangle\langle \tilde{r}_{j}\rangle}{\sigma_{i}\sigma_{j}}.
  \end{equation}

\noindent Here $\langle ... \rangle$ denotes the average over 42 trading days. We obtain a correlation matrix $C(t)$ for each time $t$ by moving the calculation window in steps of one trading day through the time series. 

All of the considered companies are classified by ten industry sectors according to the Global Industry Classification Standard (GICS) \cite{gics}. To reduce the estimation noise, we average the correlation coefficients within each sector and between different sectors, leading to $10 \times 10$ matrices. While this sector-averaged matrix is symmetric, its diagonal is not trivial. Thus, it contains $d=(10^2+10)/2=55$ independent entries. The crucial idea is to identify each averaged correlation matrix with an element in the real $d$-dimensional Euclidian space $\mathbb{R}^d$, so that a similarity, \textit{i.e.} a distance, between any two correlation matrices can be calculated. Throughout this paper all distances are measured in the Euclidean norm which we denote by $\parallel ... \parallel$. 

\begin{figure}[h]\centering
    \includegraphics[width=0.2\textwidth]{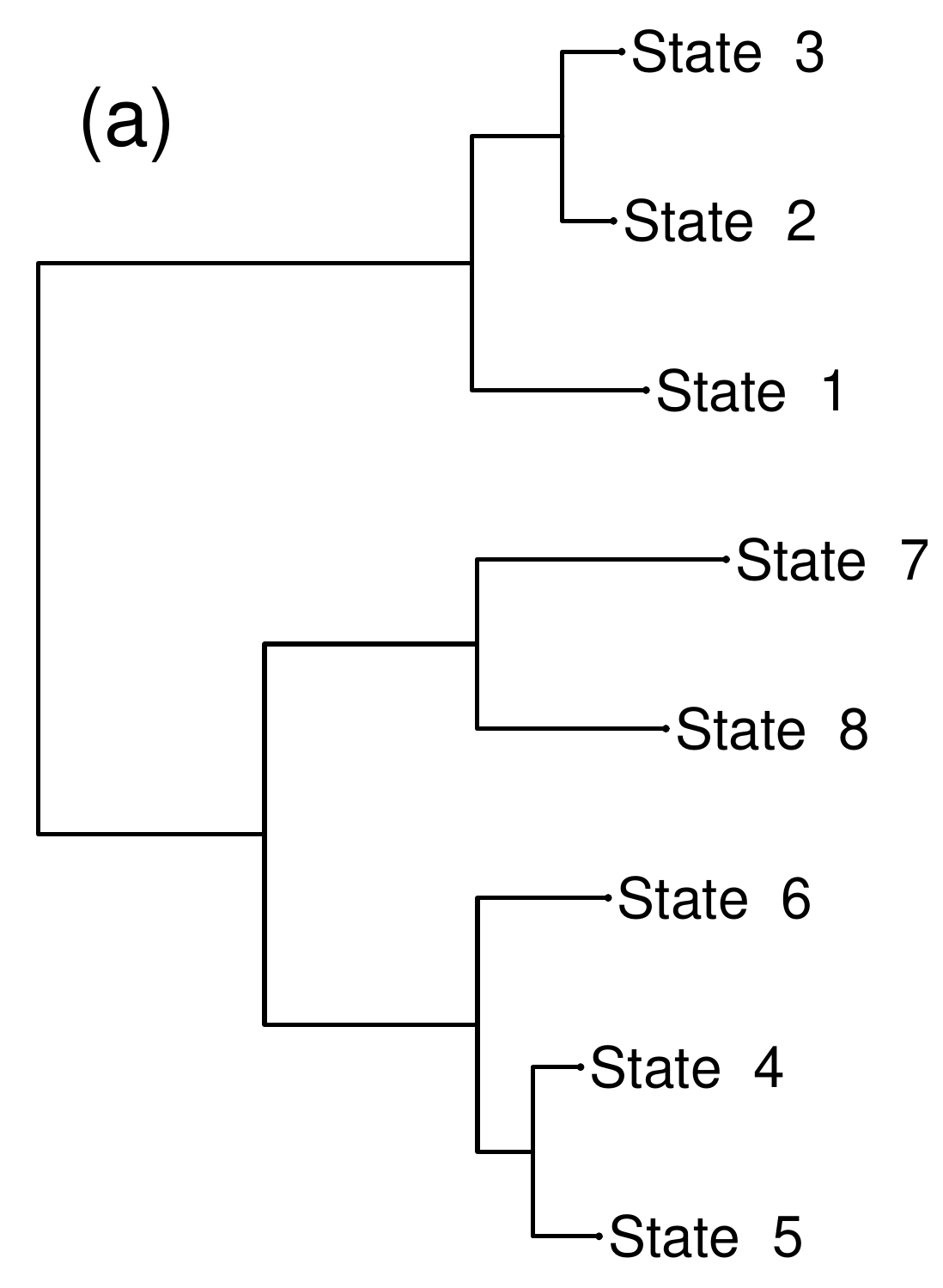}\\[2ex]
    \includegraphics[width=.49\textwidth]{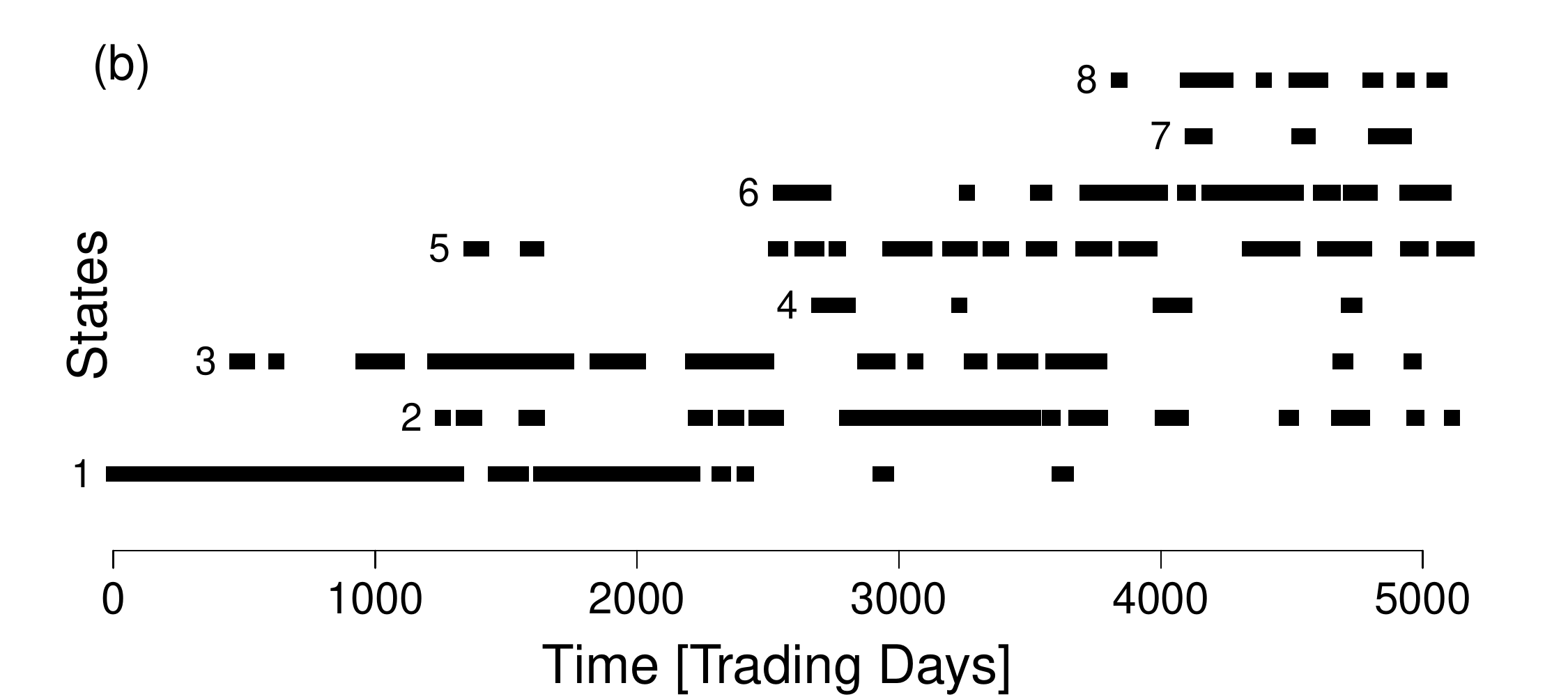}
    \caption{Clustering tree (a) and temporal evolution of the states (b).}
    \label{f.dendo+states}
  \end{figure}

As the next step we use the bisecting \textit{k}-means clustering algorithm \cite{Steinbach2000}. At the beginning of the clustering procedure all of the correlation matrices are considered as one cluster, which is then divided into two sub clusters using the \textit{k}-means algorithm with $k=2$.
This separation procedure is repeated until the size of each cluster -- in terms of the mean distance of the cluster members to the cluster center -- is smaller than a given threshold. We choose the mean distance to be smaller than $1.564$ to achieve 8 clusters as in Ref. \cite{Muennix2012}. The market is said to be in a market state $i$ at time $t$, if the corresponding correlation matrix is in the $i$th cluster. In Figure~\ref{f.dendo+states}~(a) the corresponding clustering tree is shown.
Figure~\ref{f.dendo+states}~(b) depicts the temporal evolution of the financial market. New clusters form and existing clusters vanish in the course of time. 
And while the first 1000 trading days are dominated by state 1 and only occasional jumps to state 3, we observe more frequent jumps and less stable behavior in more recent times.

\begin{figure}[h] \centering
    \includegraphics[width=.49\textwidth]{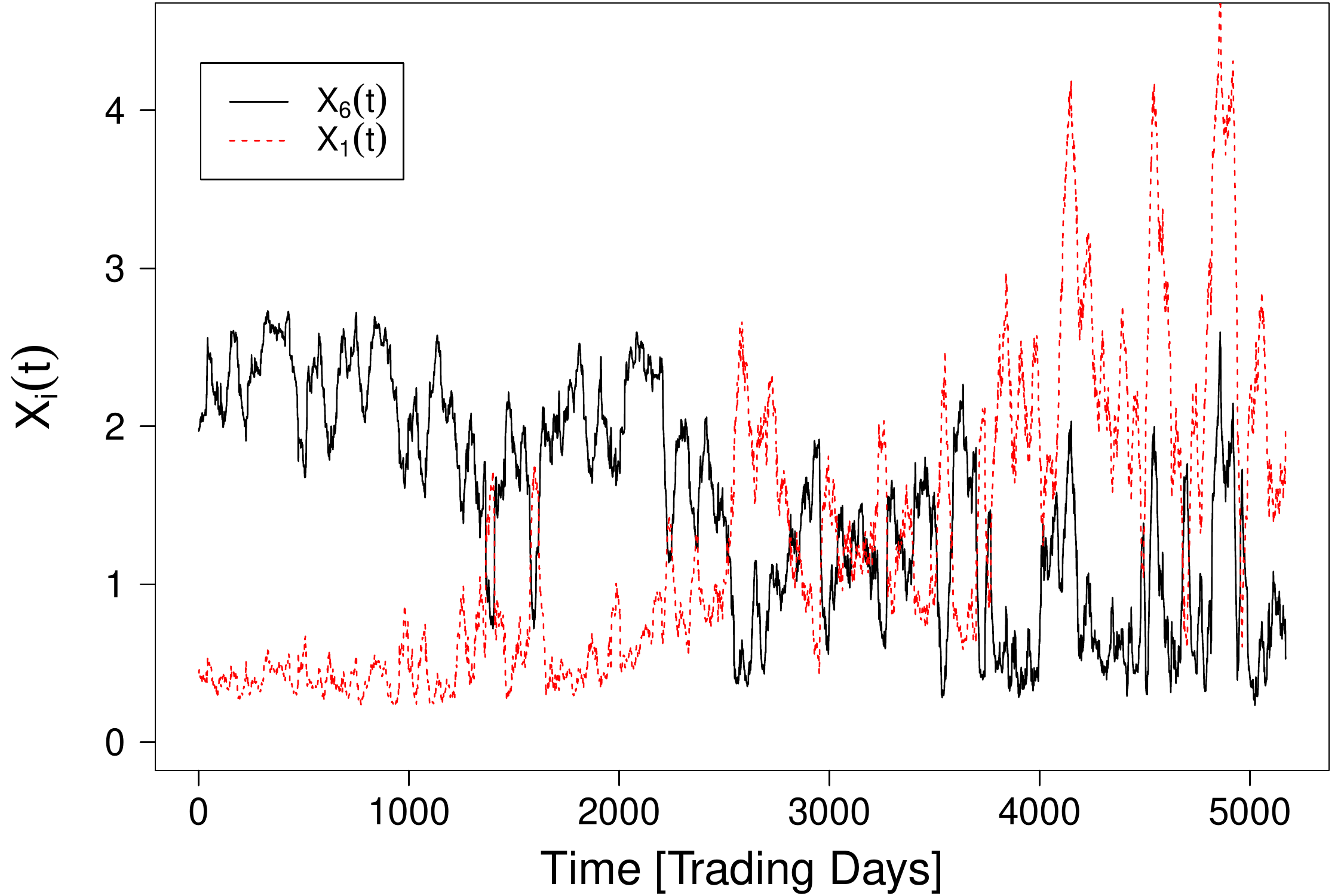}
    \caption{(Color online) Evolution of the distance (\ref{eq.timeseries}) of the system to cluster center 1 (dashed, red) and cluster center 6 (black).}
    \label{f.states}
  \end{figure}

To quantify the market situation at time $t$, the distance

\begin{equation}
X_i(t)=\parallel C(t)-\mu_i \parallel \label{eq.timeseries}
\end{equation}

\noindent of the correlation matrix $C(t)$ to the eight cluster centers $\mu_i$ ($i=1,2,...,8$) is calculated. Two of the eight resulting time series are shown in Fig.~\ref{f.states}. These time series depict the temporal evolution of the system seen from the respective cluster centers. Figure~\ref{f.states} shows that the market is in the beginning close to cluster center 1 and far away from cluster center 6. This changes over time in accordance with Fig.~\ref{f.dendo+states}~(b): cluster 6 occurs later while cluster 1 is present during the first half of the time period.

\section{Stochastic analysis}
A wide class of dynamical systems from different fields are modeled as stochastic processes and thus described by a stochastic differential equation, the Langevin equation \cite{Risken1996, Friedrich2011}

  \begin{equation}
    \dot{X}(t) = D^{(1)}(X,t) + \sqrt{D^{(2)}(X,t)} \Gamma(t)\,,
    \label{eq.Lang}
  \end{equation}

\noindent which describes the time evolution of the system variable $X$ as a sum of a deterministic function $D^{(1)}(X,t)$ and a stochastic term $D^{(2)}(X,t)\Gamma(t)$. The stochastic force $\Gamma(t)$ is Gaussian distributed with $\langle \Gamma(t) \rangle = 0$ and $\langle \Gamma(t)\Gamma(t')\rangle = 2\delta(t-t')$.
For stationary continuous Markov processes Siegert \textit{et al.} \cite{Siegert1998} and Friedrich \textit{et al.} \cite{Friedrich2000} showed that it is possible to extract drift and diffusion functions directly from measured time series using the Kramers-Moyal coefficients $M^{(n)}$, which are defined as conditional moments 

  \begin{eqnarray}
D^{(n)}(x)&=&\frac{1}{n!}\lim_{\tau\rightarrow0}\frac{M_{\tau}^{(n)}(x)}{\tau},
  \label{eq.kramers}\\
    M^{(n)}_{\tau}(x) &=& \langle (X(t+\tau) - X(t))^n \rangle \Big|_{X(t) = x} .
    \label{eq.condmom}
  \end{eqnarray}

\noindent Here $x$ denotes the value of the stochastic variable $X(t)$ at which the drift function ($n=1$) and the diffusion function ($n=2$) are evaluated. The average in Eq. (\ref{eq.condmom}) is performed over all realizations of $X(t)$ for which the condition $X(t)=x$ holds. Equation (\ref{eq.condmom}) expresses therefore (modulo the exponent $n$) the mean increment of the variable $X(t)$ after time step $\tau$ if starting at the given value $x$ at time $t$. The derivative of this mean increment with respect to $\tau$ at $\tau=0$ is equal to the value of the drift function for $n=1$ and the diffusion function for $n=2$ at $x$, as defined in Eq. (\ref{eq.kramers}). We refer to Ref. \cite{Friedrich2011} for further details. 

In the present work we estimate the drift functions on time windows of 1000 trading days, on which the estimated quantities are treated as time independent. The time dependency is obtained by comparing the estimated drift functions on a sliding time window.
Especially for small data sets the estimation of the conditional moments (\ref{eq.condmom}) might be tedious. Additionally to the original estimation procedure proposed by Siegert \textit{et al.} \cite{Siegert1998} and Friedrich \textit{et al.} \cite{Friedrich2000} we use here a kernel based regression as proposed in Ref. \cite{Lamouroux2009, Honisch2011}.

Instead of analyzing the drift function itself, it is more convenient to consider the potential function defined as the negative primitive integral
 
    \begin{equation} \label{eq.Phi}
      \Phi(x) = - \int D^{(1)}(x) \mathrm{d}x
    \end{equation}

\noindent of the drift function. The minus sign is a convention. The dynamics of the system is encoded in the shape of $\Phi(x)$: the local minima of the potential function correspond to quasi-stable equilibria, or quasi-stable fixed points, around which the system oscillates. In contrast, local maxima correspond to unstable fixed points. We note that, due to definition, potential functions are defined up to an additive constant which is here set to zero. We further note that both the drift function and the potential function have the dimension of inverse time.

\section{Optimizing the algorithm}
One drawback of clustering is that the number of clusters is in general not known \textit{a priori}. Therefore a threshold criterion is used. Furthermore, clustering is based only on geometrical properties like positions and distances between elements, \textit{i.e.} their similarity. The dynamics of the system is not involved. As a new approach we combine the clustering analysis with the stochastic analysis. The aim is to extract dynamical attributes from time series and derive stability features and quasi-stable fixed points.

From the eight time series as defined in Eq. (\ref{eq.timeseries}) we calculate the deterministic potentials $\Phi(x)_i$ ($i=1,2,...,8$) as defined in Eq. (\ref{eq.Phi}). To grasp the time evolution of the clusters, \textit{i.e.} market states, we calculate $\Phi_i(x)$ on a window of 1000 data points ($\approx$ four trading years) which we shift by 21 data points ($\approx$ one trading month), resulting in 199 deterministic potentials $\Phi_i(x)$ for each cluster $i$.

\begin{figure*}[ht] \centering
\includegraphics[width=.9\textwidth]{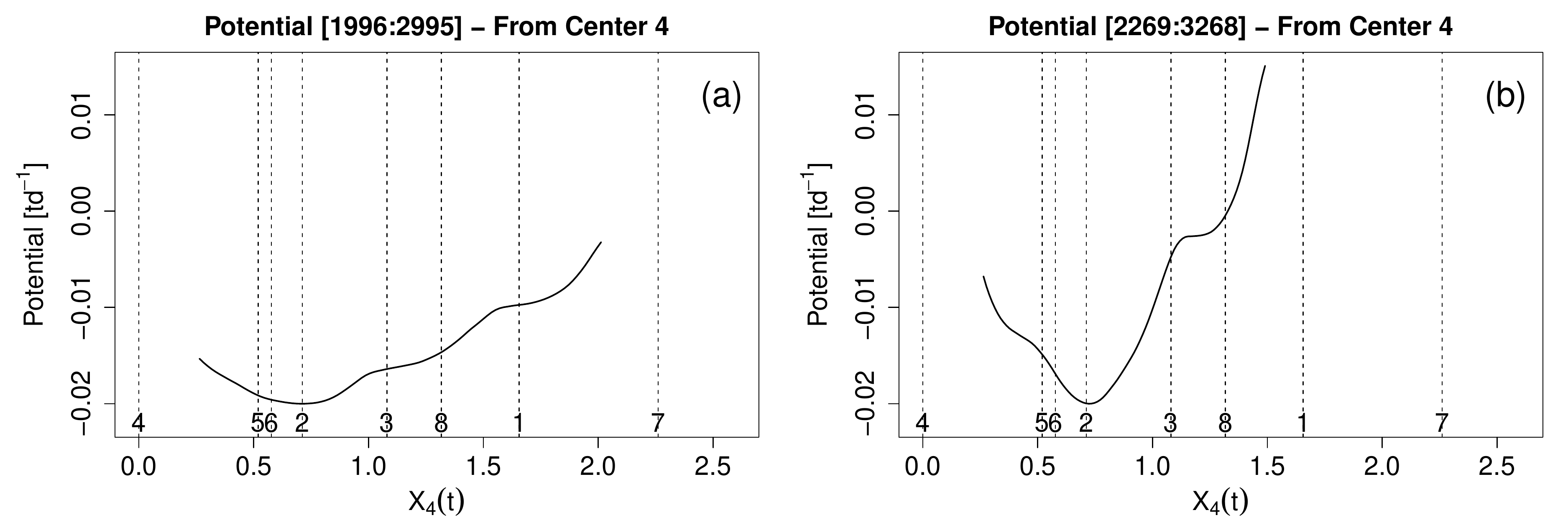}
 \includegraphics[width=\textwidth]{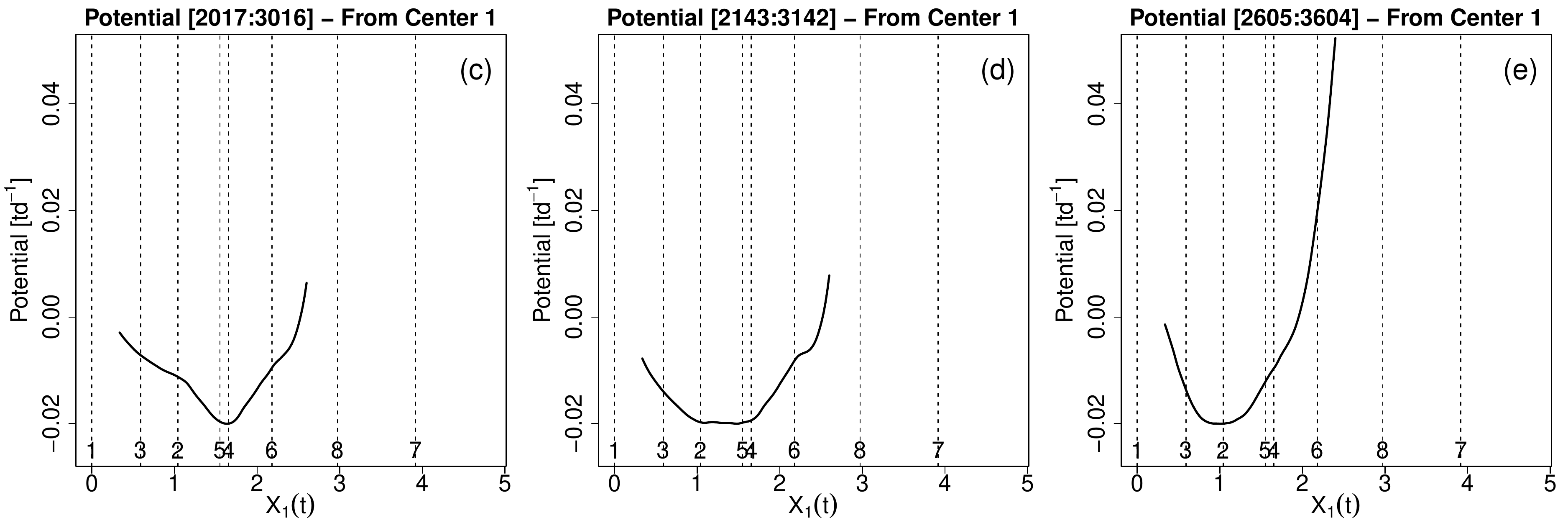}
    \caption{Potentials exibit stability of fixed points (a), (b) and transitions between system states (c) - (e). The vertical dotted lines correspond to the respective cluster centers, see Fig. \ref{f.states}. The interval in days over which the potentials were obtained is denoted by $[t_1,t_2]$.}
    \label{f.hit_centers}
  \end{figure*}

Figure~\ref{f.hit_centers} shows a sample of these potentials for different time windows. The dotted vertical lines denote the distance to the cluster centers which are labeled at the abscissa. 
Each potential in the five figures shows a clear minimum. Hence, the market dynamics expressed by the correlation matrices performs a noisy dynamics around the attractive fixed point, which is defined by the minimum. Most interestingly the position of the minima of these potentials coincide quite well with the distances to the cluster centers obtained from the cluster analysis. Additionally to the positions of the fixed points of the system, the potentials provide information about the stability of the market in the analyzed time window. Potential functions with more than only one clear local minimum, e.g. Fig.~\ref{f.hit_centers}~(a)~and~(b), reflect an unstable dynamics. In contrast an isolated and deep minimum as in Fig.~\ref{f.hit_centers}~(e), corresponds to a stable dynamics.

In Fig.~\ref{f.hit_centers}~(c)~-~(e) a transition from state 4 and 5, which are very  close, to state 2 is shown. The quasi-stable fixed point of the potential function changes its position in time and moves closer to the center of the first cluster. We note that in the intermediate state, as shown in Fig.~\ref{f.hit_centers}~(d), the width of the potential function is increased. Instead of a clear local minimum it has a rather flat plateau. The time evolution of the market reflected in the position change of the local minimum of the potential function is thus taken as a state transition. The ability of our approach to describe multi-stable and transitional behavior has also been shown by M\"ucke \textit{et al.} \cite{Muecke2014} in the context of wind energy. Stability of market states as well as state transitions are studied in detail in Stepanov \textit{et al.} \cite{msdyn}.

\section{Merging of the clusters}
\begin{figure}\centering
    \includegraphics[width=.45\textwidth]{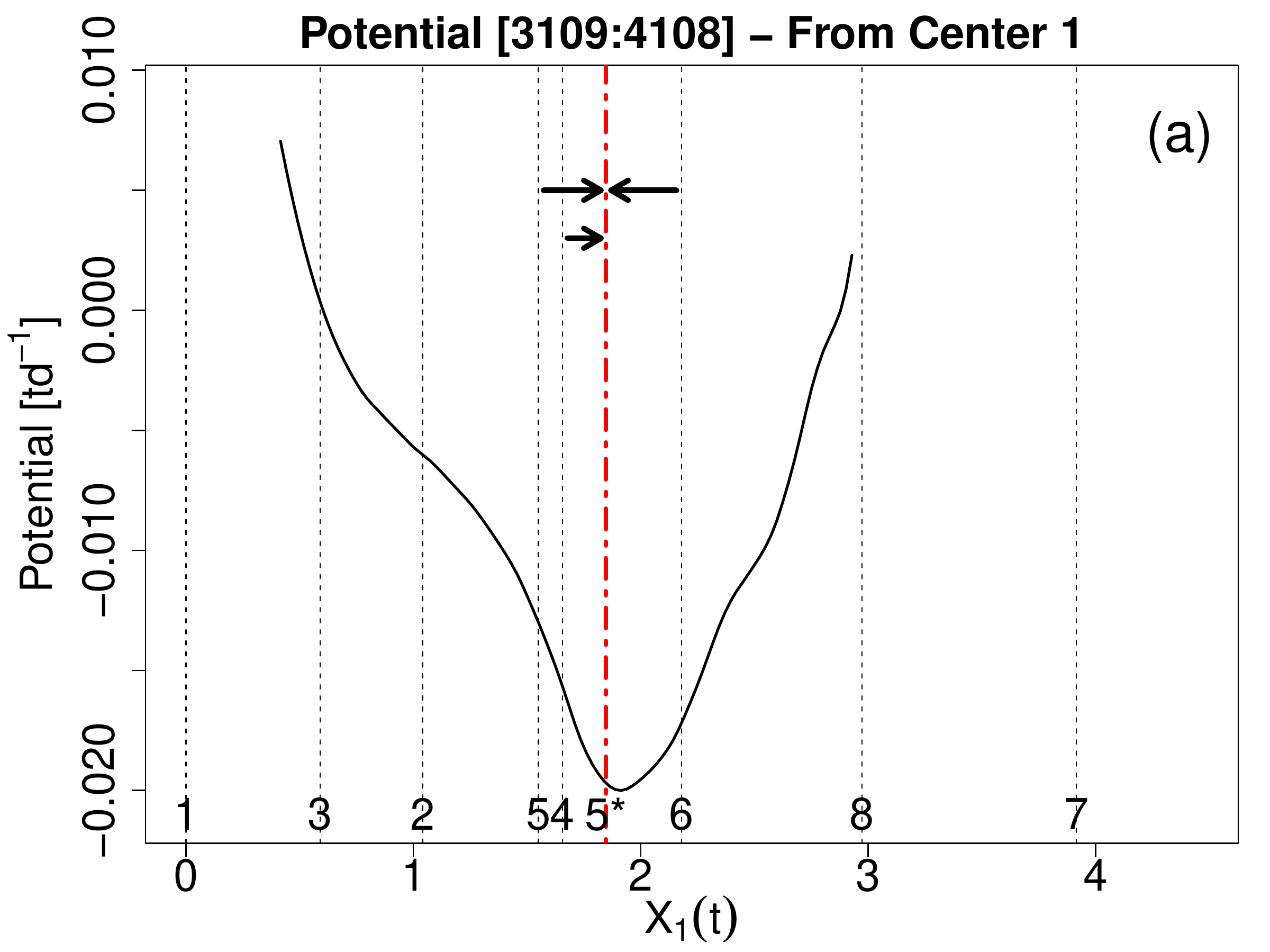}
    \includegraphics[width=.45\textwidth]{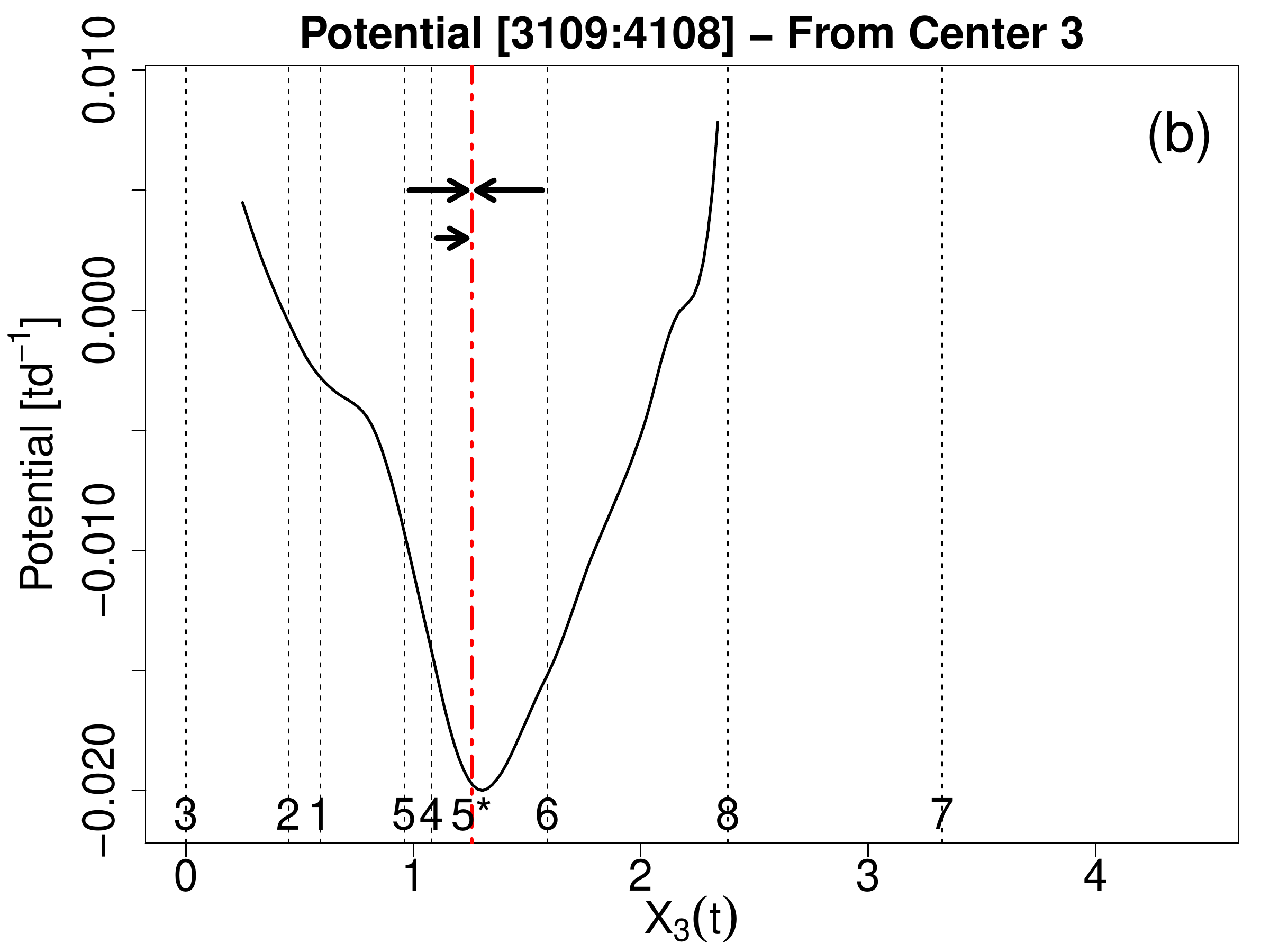}
    \caption{(Color online) Result of merging states 4, 5 and 6 to match the minimum of the potential seen from state 1 (a) and state 3 (b).}
    \label{f.merge}
  \end{figure}

Besides the cases where the positions of the minima of the potentials coincide clearly with the distances to cluster centers, there are also less clear situations as shown in Fig.~\ref{f.merge}. In the considered time window $I=[3109:4108]$ (expressed in td) the market is switching between the states 2, 3, 4, 5 and 6. The clustering procedure as described above doesn't take the dynamics of the market into account. All of the correlation matrices are clustered at the same time without any information about how the market is jumping between the correlation matrices. The elements of a cluster are chosen in such a way that the sum of the distances to its cluster center is minimized.
 
The stochastic analysis of the market dynamics seen from cluster center 1 as well as from cluster center 3, shows that the potential function exhibits a pronounced minimum between the distances to the cluster centers 5, 4 and 6 as shown in Fig.~\ref{f.merge}. This fact is now taken as a criterion to change the cluster analysis in a way that the clusters 4, 5 and 6 will be combined to a new cluster 5*. That this merging of clusters is natural is indicated by the clustering tree of Figure~\ref{f.dendo+states}~(a), as the three merged clusters arise from the split of the same ancestor cluster. If the threshold is chosen small enough, this cluster would not have been split. 

\begin{figure}
    \centerline{\includegraphics[width=.98\columnwidth]{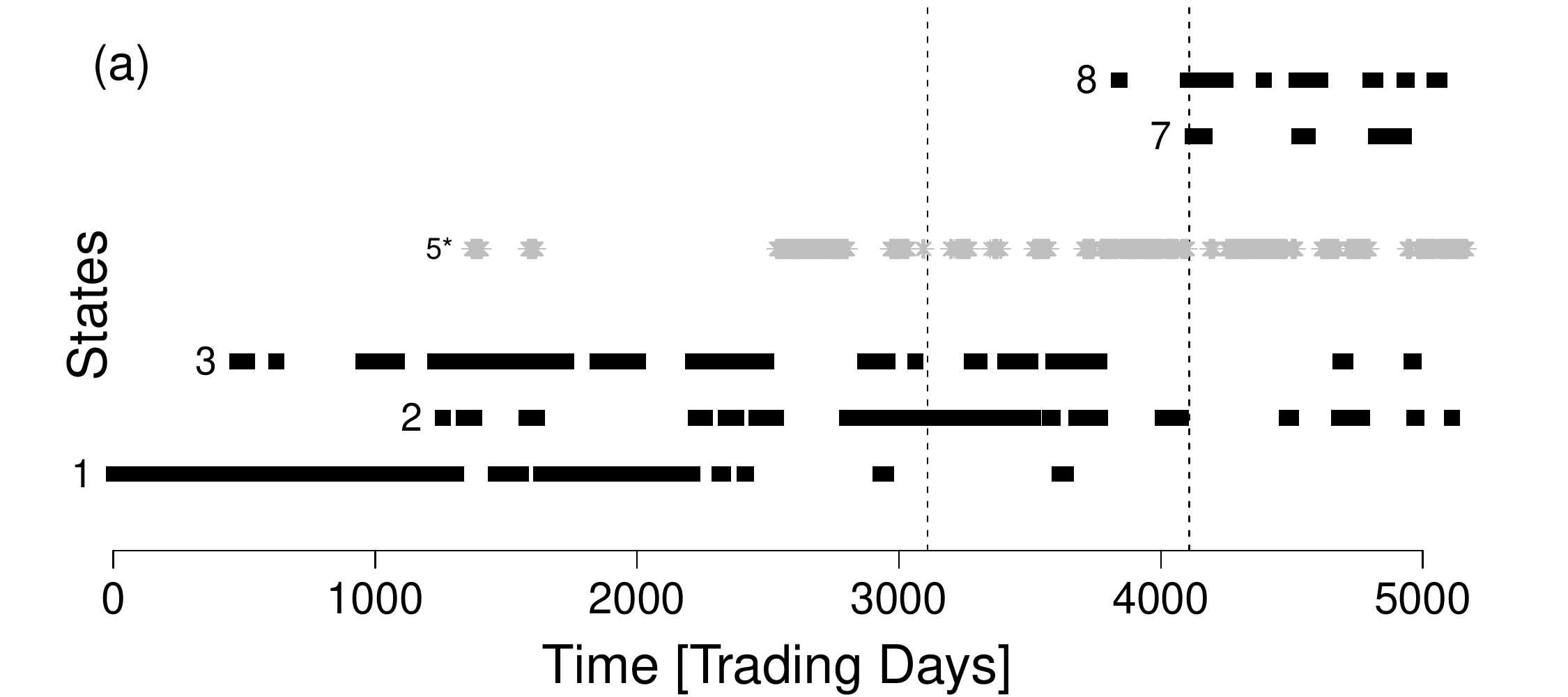}}
        \centerline{\includegraphics[width=.98\columnwidth]{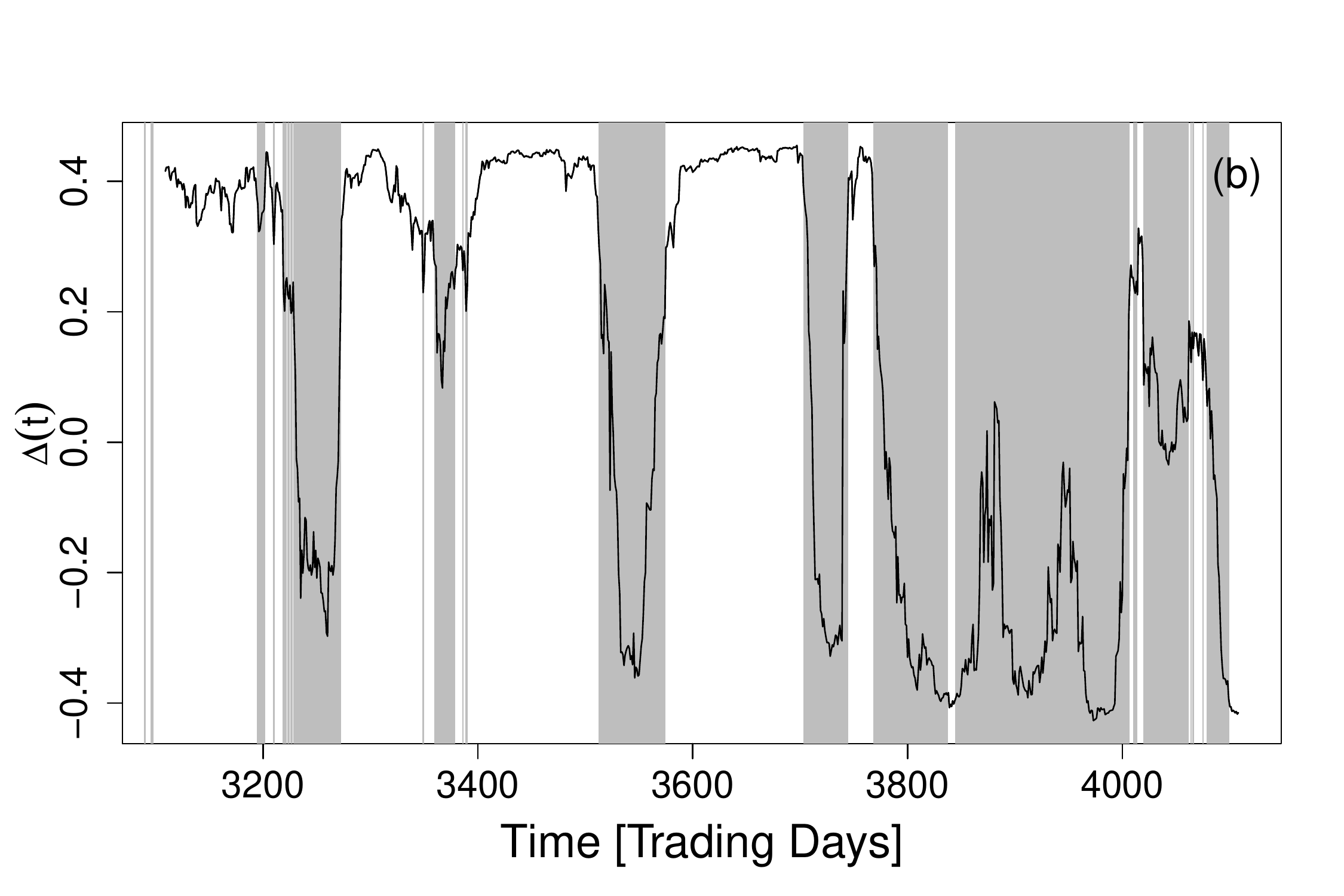}} 
        \caption{Temporal evolution of the remaining six states (top). The merged market state is denoted by 5*. The lower plot shows the time evolution of $\Delta(t)$ as defined in Eq.~(\ref{eq.d}). The time points in which the market occupies the merged state 5* are marked by the vertical grey lines.}
    \label{f.evo_states_6}
  \end{figure}

For reference we show in Fig.~\ref{f.evo_states_6} the temporal evolution of the states after merging clusters 4, 5 and 6 to cluster 5*. 
The distance to the center of the merged cluster as seen from clusters 1 and 3, respectively, coincides remarkably with the position of the local minimum of the potential as marked by the broken line (dash point) in Fig.~\ref{f.merge}. We thus conclude that while for the chosen threshold the clusters, \textit{i.e.} market states 4, 5 and 6 are geometrically distinct, together these states form a single quasi-stationary state. While being in this state, the system fluctuates around the center of the cluster 5*, which is a fixed point of the system. We note that the mean correlation matrix

\begin{equation} \label{eq.meanCorr}
\bar{C}=\frac{1}{N_I}\sum_{t\in I}{} C(t) 
\end{equation}

\noindent of the analyzed time period $I$ differs clearly from the center of the merged cluster $\mu_{5^*}$ with $\parallel \mu_{5^*} - \bar{C} \parallel=0.55$. Here $N_I$ denotes the length of $I$. Furthermore the distances to the cluster center 1 $\parallel \mu_{1} - \bar{C} \parallel=1.42$ and the cluster center 3 $\parallel \mu_{3} - \bar{C} \parallel=0.82$ don't match the positions of the minimum of the potential as shown in Fig.~\ref{f.merge}.

\section{Identifying high-dimensional fixed points} 
In the previous section we used clustering and stochastic analysis of the one-dimensional time series to obtain fixed points of the system. We identified the positions of the local minima of the potential functions with the distances to the quasi-stationary fixed points. In this section we propose a method to identify quasi-stationary fixed points by an optimization problem. 

We recall that the mean correlation matrix (\ref{eq.meanCorr}) minimizes the sum of distances

\begin{equation} \label{eq.SumDist}
\sum_{t\in I}{}\parallel \Lambda-C(t)\parallel
\end{equation}

\noindent of $C(t)$ in $I$ to a fixed correlation matrix $\Lambda$. For high-dimensional empirical data, distinct subspaces (principal components) may exist along which data points are preferably distributed \cite{NLDR,Hotelling,msdyn}. Therefore we require the fixed points of the system to be elements of these subspaces. As the sum of two elements of a given principal component is not necessarily an element of this component anymore, the calculation of the average (\ref{eq.meanCorr}) does not neccessarily yield the fixed point of the system. We gave an example of this in the previous section. The calculated average $\bar{C}$ does not match the position of the fixed point found by stochastic analysis.

Here we propose to identify system fixed points as the elements of the preferred subspaces which are the most similar to all $C(t)$ in a given time interval $I$, i.e. they minimize the sum of distances (\ref{eq.SumDist}). In absence of any preferable subspaces we recover the mean correlation matrix $\bar{C}$. This idea is similar to the definition of the geodetic curves on an arbitrary manifold which minimize the distance between any two points and are not necessarily straight lines any more.

As an application we consider the data sample from the previous section. As shown in Fig.~\ref{f.merge} the stochastic analysis in the time interval $I=[3109:4108]$ (expressed in trading days) shows a deep minimum at $X_0 \approx 1.9$ as seen from the center of the first market state $\mu_1$. The data points are therefore distributed approximately around $\mu_1$. The set of all possible fixed points is restricted to the points on the $(d-1)$-dimensional hypersphere of radius $X_0$ around $\mu_1$. To find the empirical fixed point for this setting we minimize (\ref{eq.SumDist}) under the condition 

\begin{equation}
\parallel \mu_1 - \Lambda \parallel \equiv X_0 \approx 1.9,
\end{equation}

\noindent which we solve by the method of Lagrange multipliers. We note that for the Euclidian norm the problem always has a unique minimum and maximum, unless the distance (\ref{eq.timeseries}) is constant. 

The empirical solution of the problem, denoted by $C_0$, differs slightly from the center of the merged cluster defined in the previous section, $\parallel C_0-\mu_{5^*}\parallel \approx 0.35$. This is because the market does not exclusively stay in the merged state 5* during the analyzed time period $I$, as observed in Fig.~\ref{f.evo_states_6}~(a). 

We now quantify the deviation of the obtained fixed point $C_0$ from the averaged correlation matrix $\bar{C}$ by looking at the difference 

\begin{equation}
\Delta(t)=\parallel C(t)-C_0 \parallel - \parallel C(t)-\bar{C} \parallel
\label{eq.d}
\end{equation} 

\noindent of the distances from $C(t)$ to $C_0$ and $\bar{C}$, respectively. The market is closer to $C_0$ than to $\bar{C}$ whenever $\Delta(t) <0$ holds. 

The time series $\Delta(t)$ is shown in Fig.~\ref{f.evo_states_6}~(b). It is switching between positive and negative values.
The grey background highlights the time points $t$ at which the market occupies the merged state 5*, which remarkably coincidences with the negative values of $\Delta(t)$. We almost exactly quantify the merged market states during which $\Delta(t)$ falls rapidly down from positive values. Not only is the market state identified but also the dynamics of the market within the state. As seen from Fig.~\ref{f.evo_states_6}~(b), the values of $\Delta(t)$ continuously decrease and then increase again while the market occupies the merged state. In contrast $\Delta(t)$ is fluctuating around a positive value while the market is not in the merged state.

As a consistency check, we applied the same algorithm with the center of the third cluster $\mu_3$, as well the overall mean correlation matrix as reference points. In all cases we obtained the same fixed point $C_0$.

\section{Conclusions}
The combination of the cluster analysis and the stochastic process analysis allows to characterize the dynamics of a dynamical system as a noise process between different quasi-stationary states. For financial markets these market states are defined in terms of correlation matrices which reflect the dependence structure of the stock market. Especially, while being in a given market state, the market is fluctuating around the center of the corresponding cluster. A threshold criterion can produce geometrically distinct states, which are a single quasi-stationary state of the system. The deviation of the market situation at time $t$ from individual market states is reflected in the distance between the correlation matrix $C(t)$ and the respective cluster centers. This distance is taken as the new low-dimensional order parameter of the complex system. The stochastic analysis provides evidence of how the market dynamics is guided by a changing potential landscape with temporally changing stability of the market states. Emerging new quasi-stable states are found. In this way we present a method with which the dynamics of a high-dimensional complex system is projected to low-dimensional collective dynamics. Furthermore we address the high dimensionality of the data set by an optimization problem. This problem is well defined and is robust against the choice of reference points. We obtain high-dimensional quasi-stable fixed points of financial markets explicitly and see good chances to apply this method also to other complex states.

\acknowledgments
Philip Rinn thanks David Bastine for fruitful discussions.


\end{document}